\documentclass[aps,showpacs,reprint,amsmath,amssymb,groupaddresses,superscriptaddress]{revtex4-1}

\usepackage{graphicx}
\usepackage{dcolumn}
\usepackage{bm}

\begin{document}


\title{Optically Induced Nuclear Spin Polarization in the Quantum Hall Regime: \\
The Effect of Electron Spin Polarization through Exciton and Trion Excitations}



\author{K. Akiba}
\altaffiliation{Present affiliation: Department of Applied Physics, Tokyo University of Agriculture and Technology}
\affiliation{Department of Physics, Tohoku University, Sendai, 980-8578, Japan}
\affiliation{JST, ERATO Nuclear Spin Electronics Project, Sendai, 980-8578, Japan}

\author{S. Kanasugi}
\affiliation{Department of Physics, Tohoku University, Sendai, 980-8578, Japan}

\author{T. Yuge}
\altaffiliation{Present affiliation: Department of Physics, Shizuoka University}
\affiliation{Department of Physics, Osaka University, Machikaneyama-Cho, Toyonaka, 560-0043, Japan}

\author{K. Nagase}
\altaffiliation{Present affiliation: Graduate School in Spintronics, Tohoku University}
\affiliation{Department of Physics, Tohoku University, Sendai, 980-8578, Japan}
\affiliation{JST, ERATO Nuclear Spin Electronics Project, Sendai, 980-8578, Japan}

\author{Y. Hirayama}
\affiliation{Department of Physics, Tohoku University, Sendai, 980-8578, Japan}
\affiliation{JST, ERATO Nuclear Spin Electronics Project, Sendai, 980-8578, Japan}
\affiliation{WPI-AIMR, Tohoku University, Sendai, 980-0812, Japan}


\date{\today}

\begin{abstract}
We study nuclear spin polarization in the quantum Hall regime 
through the optically pumped electron spin polarization in the lowest Landau level. 
The nuclear spin polarization is measured as a nuclear magnetic field $B_N$
by means of the sensitive resistive detection. 
We find the dependence of $B_N$ on filling factor unmonotonous. 
The comprehensive measurements of $B_N$ 
with the help of the circularly polarized photoluminescence measurements 
indicate the participation of the photo-excited complexes 
i.e., the exciton and trion (charged exciton),  
in nuclear spin polarization. 
On the basis of a novel estimation method of the equilibrium electron spin polarization, 
we analyze the experimental data 
and conclude that the filling factor dependence of $B_N$ is understood by 
the effect of electron spin polarization through excitons and trions. 
\end{abstract}




\maketitle

The coupling between electron and nuclear spins 
through the contact hyperfine interaction 
realizes the dynamic nuclear spin polarization 
and the detection of a small ensemble of nuclear spins. 
This allows us to perform nuclear magnetic resonance (NMR) in a microscopic region 
through electrical or optical manipulation of electron spins \cite{SpinPhysics}.
This new type of NMR technique 
is a powerful tool to probe electronic properties
and also has a potential 
to implement quantum information processing by using nuclear spins as qubits.  
Indeed, its intriguing electronic properties have been revealed in the quantum Hall system \cite{Smet, Kumada, Muraki1, Muraki2, Vitkalov1}, 
and multiple quantum coherences of nuclear spins have been controlled in a nanometer-scale region \cite{Yusa}.
In these experiments, 
electrical pumping and resistive detection of nuclear spins play an important role,  
while optical pumping of nuclear spins has also been achieved \cite{Barrett, OPRD1, Vitkalov2, Kuku, Davies}. 
The condition for the electrical pumping is restricted to 
a special electronic state such as spin phase transition in $2/3$ fractional quantum Hall state
and this limits the application of the NMR technique. 
However, the optical pumping does not bring about this restriction.  
When one combines electrical and optical means, 
the fascinating electronic states in the quantum Hall system will be widely investigated
and rich quantum information processing can be demonstrated. 

The research on the optical pumping conditions for the dynamical nuclear polarization 
has been performed in the quantum Hall regime \cite{Barrett,OPRD2}. 
However, how the quantum Hall electronic states affect its polarization 
has not been fully investigated. 
In this Letter, we study the dependence of the optically induced nuclear spin polarization on the electric state in the quantum Hall regime 
i.e., the Landau level filling factor $\nu$. 
We find a correlation between the nuclear polarization and the photoluminescence (PL). 
Our experimental data are analyzed 
by use of the estimation of electron spin polarization that we constructed.  
We understand the $\nu$-dependence of the optical nuclear polarization 
as the effect of the electron spin polarization through excitons and trions in the quantum Hall regime. 

Experiments were carried out on a single 18-nm GaAs/Al$_{0.33}$Ga$_{0.67}$As quantum well 
with single-side doping, which was processed to a 100-$\mu$m-long and 30-$\mu$m-wide Hall bar. 
The electron density $n_s$ of the 2-dimensional electron system (2DES) can be tuned by applying a voltage to the $n$-type GaAs substrate (back gate).
The sample was cooled in a cryogen free $^3$He refrigerator down to 0.3~K
and pumped by a mode-locked Ti:sapphire laser (pulse width: $\sim 2$ ps, pulse repetition: 76 MHz).
The electron mobility is 185 m$^2$/(Vs) for $n_s=1.2\times 10^{15}$ m$^{-2}$. 
A laser beam (diameter: 230 $\mu$m) irradiated the whole Hall bar structure
through an optical window on the bottom of the cryostat. 
The propagation direction of the laser beam was parallel to the external magnetic field $B=7.15$ T, 
which was perpendicular to the quantum well. 
We can vary $\nu=h n_s/ (e B)$ using the back gate, 
where $h$ is the Planck constant and $e$ is the elementary charge. 

The optical pumping was performed as follows.
First, the nuclear spin polarization was fully destroyed 
by setting the electronic state to the skyrmion region \cite{Smet}. 
Second,  
right or left circularly polarized light ($\sigma^+$ or $\sigma^-$) illuminated the sample, 
where the electronic state was set to $\nu$ during illumination. 
The pumping time was 250~s, which was long enough to saturate the optical nuclear polarization. 
The pumping photon energy $E_{\textrm{laser}}$ and the average power density $P$ 
are specified below.
The laser illumination increased the temperature of the sample holder up to 0.4~K
and also disguised the sample resistance. 
Third, after optical pumping, $\nu$ was set to 1 for 70~s 
so that the resistance returned to the value before illumination, 
where the relaxation of nuclear polarization at $\nu=1$ is the smallest within the available $\nu$. 
This relaxation time was over $1.6\times10^{3}$~s.
Therefore, the nuclear spin polarization generated by optical pumping was not destroyed 
during the waiting time at $\nu=1$.

The optically induced nuclear polarization was measured 
by the resistive detection method 
using a peak shift of the spin phase transition at $\nu=2/3$ \cite{OPRD1, OPRD2}. 
This method is a highly sensitive detection of nuclear polarization.  
Here, we recorded the nuclear magnetic field $B_N$ 
induced by the nuclear polarization 
of the relevant three nuclides ($^{69}$Ga, $^{71}$Ga and $^{75}$As) 
at the electric-current-flowing region. 
We used a standard low-frequency (83~Hz) and low-current (30~nA) 
lock-in technique to measure the resistance. 
Thus, the resistive detection method we used 
(see our previous paper \cite{OPRD2} for the experimental details) enables to 
probe only a small ensemble of nuclear spins interacting with the 2DES.

\begin{figure}
\includegraphics[width=0.75\columnwidth]{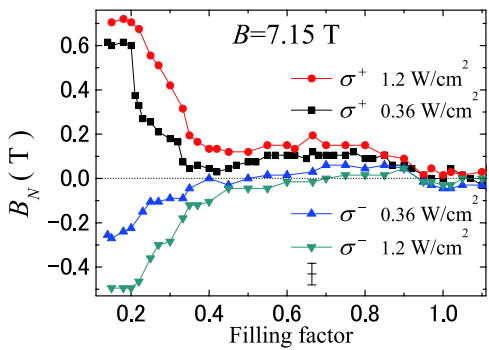}%
\caption{(Color online) The $\nu$-dependence of optical nuclear spin polarization at $B=7.15T$.
The error bar shows typical errors in $B_N$.
\label{fig:FillingDep.eps}}
\end{figure}

Figure~\ref{fig:FillingDep.eps} shows the $\nu$-dependence of optically induced $B_N$ 
for $\sigma^{+}\ (\sigma^{-})$ excitation with $E_{\textrm{laser}}=1.5328\ (1.5321)$~eV. 
Here, $\sigma^{\pm}$ excitation is associated with 
the interband transition from a heavy hole band with angular momentum $J_z=\mp3/2$ 
to the lowest electron Landau level with spin $S_z=\mp1/2$.
$B_N$ induced by $\sigma^{+}$ and $\sigma^{-}$ excitations 
show the opposite direction  
because the conduction electrons with down and up spins are created by 
$\sigma^{+}$ and $\sigma^{-}$ excitations, respectively \cite{OPRD2}.  
When $\nu$ increases from the lower-side, the magnitude of $B_N$ decreases for both excitations.
In the $\nu$-range from 0.4 to 0.9,
relatively small values of $B_N$ are observed for $\sigma^{+}$ 
and the apparent nuclear spin polarization is not observed for $\sigma^{-}$.
Around $\nu=1$, nuclear spins are not polarized for either excitations. 
The stronger excitation power exhibits the larger magnitude of $B_N$, 
which is explained by the increased pumping rate. 
However, the non-monotonic behavior of the $\nu$-dependence remains unchanged. 
There are three regions for optical nuclear spin polarization in the lowest Landau level: 
(I) $\nu<0.4$, (II) $0.4<\nu<0.9$ and (III) $\nu>0.9$.

\begin{figure}
\includegraphics[width=0.80\columnwidth]{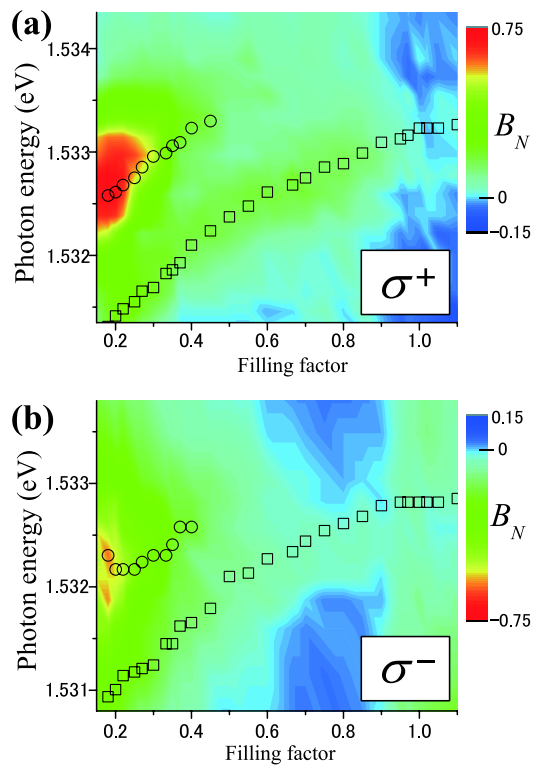}%
\caption{(Color online) The 2D color map of $B_N$ 
for (a) $\sigma^{+}$ and (b) $\sigma^{-}$ excitations. 
The scale of color bars is linear. 
The directions of the bars for (a) and (b) are reversed for clarity. 
The circles (squares) show the photoluminescence peak positions of the triplet (singlet) trion.
\label{fig:2Dmap6.eps}}
\end{figure}

To investigate these behaviors, 
we measured $\nu$-dependence of optically induced $B_N$ 
by changing $E_{\textrm{laser}}$ with $P=1.2$~W/cm$^2$. 
The optical pumping rate is expected to depend on the photon absorption rate, 
and $\nu$-dependence with constant $E_{\textrm{laser}}$ should be modified 
when the absorption spectrum is varied by changing $\nu$. 
The optical transitions in the quantum Hall system (both absorption and luminescence) 
are determined by 
the strong Coulomb interaction between the valence hole and the surrounding electrons, 
resulting in the existence of bound electron-hole complexes, 
e.g., neutral and charged (trions) excitons in the lowest Landau level \cite{BarJoseph}. 
The configuration of our sample is not suitable for absorption measurements. 
Although the peak positions of the absorption and luminescence 
are not completely coincident, 
the luminescence peak can be used as the indicator of the absorption peak  
due to the relatively small energy difference \cite{OSdarktriplet}.
Therefore, we also measured the circularly polarized PL  
with a spectral resolution better than 0.2~meV, 
where we used the linearly polarized light \cite{ReasonLP}
with the excitation energy of 1.58~eV and the power density of 1.2~W/cm$^2$.

Figure~\ref{fig:2Dmap6.eps} (a) ((b)) shows the color map of $B_N$
for $\sigma^{+}$ ($\sigma^{-}$) excitation. 
The transverse and longitudinal axes indicate $\nu$ and photon energy, respectively. 
The $\sigma^{+}$ ($\sigma^{-}$) PL peak positions 
of triplet (circles) and singlet (squares) trions are overlaid in (a) ((b)), 
where the peaks were assigned by the $B$- and $n_s$-developments \cite{YusaPE} 
and the triplet and singlet mean the spin alignments of two electrons in the trion. 
The increase of the peak energies as $\nu$ increases 
is understood by the quantum confined Stark effect 
because we controlled $\nu$ using the gate voltage. 
Taking into consideration the spectral width of the pumping laser 
(full with at half maximum $\sim$0.6~meV), 
we find that the nuclear polarization basically occurs at the PL peak positions. 
The nuclear polarization at the triplet trion peak is larger 
than that at the singlet trion peak. 
Thus, in terms of the PL peak as the indicator of the absorption peak 
(the details will be discussed in the next paragraph.), 
we observe the correlation between the nuclear polarization and the photoexcited complex absorption.
This accounts for the difference between regions (I) and (II) in Fig.~\ref{fig:FillingDep.eps}.

We here consider what information for the photon absorption is elicited from the observed PL
since the absorption is important to polarize nuclear spins as mentioned above. 
Although we assigned the upper PL peak to the triplet trion, 
the neutral exciton peak is expected to be merged into (or have slightly higher energy than) 
the triplet trion peak \cite{OSdarktriplet, YusaPE} under our experimental conditions. 
The neutral exciton has greater oscillator strength than the triplet trion 
in the absorption measurement \cite{OSdarktriplet} 
and in the numerical calculations in high $B$-field \cite{Wojs}. 
Therefore, we can attribute the nuclear polarization at the triplet PL peak to 
the absorption of the neutral exciton \cite{TTcontPL}.
In contrast, we consider the nuclear polarization at the singlet PL peak 
as the consequence of the absorption of the singlet trion \cite{QDtrion}. 
Indeed, in the absorption experiment (under conditions similar to ours) 
performed by Groshaus {\it{et al.}} \cite{Groshaus}, 
two peaks were assigned to the neutral exciton $X$ and the singlet trion $T$ \cite{NoteA}.

Next, we discuss how the optically excited complexes affect the nuclear spin polarization.
Our experimental results indicate that 
the photo-excitation of $X$ leads to higher nuclear polarization than that of $T$. 
The photon absorption rate is proportional to the number of the injected electron spins, 
which subsequently polarizes the nuclear spin.
$X$ in high $B$-field or at low $n_s$ 
is expected to have larger absorption than $T$ \cite{OSdarktriplet, Wojs}. 
This can explain our results simply. 
However, the absorption measurement does not always show such behavior 
under the experimental conditions similar to ours \cite{Groshaus}. 
To polarize nuclear spins, primarily, 
the electron spin polarization under optical pumping $\langle S_z \rangle$ 
is more crucial than the number of injected electron spins.  
The effective nuclear magnetic field after long pumping time is given by 
$B_N=-A\,(\langle S_z \rangle - \langle S_z \rangle_{\rm{eq}}) \label{BN}, $
where $A(>0)$ is a constant 
and $\langle S_z \rangle_{\rm{eq}}$ is the equilibrium electron spin polarization \cite{OPRD2,Note1}.
This fact and the experimental results indicate that 
the photo-excitation of $X$ generates higher $\langle S_z \rangle$ than that of $T$ 
in the quantum Hall regime. 
Indeed, this can be expected from the study of optical spin pumping 
in the II-VI quantum well \cite{Astakhov}. 
Moreover, there is a possibility that $X$ directly and indirectly polarizes the nuclear spins,   
because the electron in $X$ has $s$-type symmetry, 
which considerably contributes to the contact hyperfine interaction, 
and because $X$ forms $T$ by capturing the resident electron.

\begin{figure}
\includegraphics[width=0.65\columnwidth]{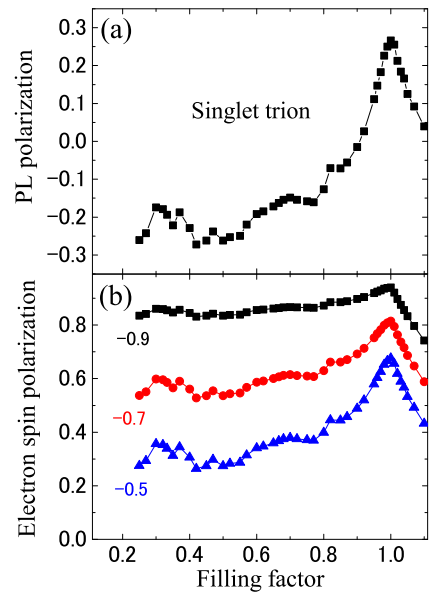}%
\caption{(Color online) 
(a) The PL polarization obtained from the singlet-trion peak intensities with the 1.2~W/cm$^2$ linearly polarized excitation. 
(b) The electron spin polarizations estimated from (a) with $P_h=-0.9$, $-0.7$ and $-0.5$.
\label{fig:PLpolarization.eps}}
\end{figure}

We also take into consideration $\langle S_z \rangle_{\rm{eq}}$ 
to understand the $\nu$-dependence of $B_N$. 
To know the electron spin polarization $P_e$, 
the optical dichroism calculated from the trion absorption is available \cite{Groshaus}. 
Although we cannot measure the absorption of our sample,  
the PL polarization $P_{L}$ has a contribution of $P_e$ 
and has been utilized to extract the $P_e$ characteristics \cite{Pusep1, Pusep2}. 
We develop the $P_e$ estimation from $P_{L}$. 
The $\sigma^+ (\sigma^-)$ PL intensity $I_{\sigma^{+} (\sigma^{-})}$ is proportional to 
the number of photo-excited particle multiplied by its oscillator strength. 
We define $P_{L}$ as $(I_{\sigma^{+}}-I_{\sigma^{-}})/(I_{\sigma^{+}}+I_{\sigma^{-}})$ 
and here consider this formulation for $T$. 
Since the photo-created electron needs to pair with an opposite spin, 
the oscillator strength of each $T$ can be modeled 
as proportional to the number of unpaired electrons with opposite spin \cite{Groshaus}. 
Consequently, the calculation of $P_{L}$ for $T$ gives $P_e=(P_L-P_h)/(1-P_L\,P_h)$ for $\nu \le 1$ 
and $P_e=(2-\nu)/\nu \cdot (P_L-P_h)/(1-P_L\,P_h)$ for $\nu > 1$, 
where $P_h$ is the hole polarization due to the singlet nature of $T$ \cite{Suppl}. 
While $P_h$ increases with $B$ \cite{Kukushkin}, 
our experiments were performed under constant $B$, 
and $P_h$ should be constant. 
Figure~\ref{fig:PLpolarization.eps} (a) and (b) respectively show the measured $P_{L}$ for $T$
and the $P_e$ calculated with the $P_h$ values of $-0.9$, $-0.7$ and $-0.5$. 
Since the optical pumping was performed with the strong illumination, 
the $P_e$ obtained here (under the strong photo-excitation) 
is treated as $\langle S_z \rangle_{\rm{eq}}$ \cite{NoteStrong}. 
Thus, we obtain the trend of $\langle S_z \rangle_{\rm{eq}}$ 
although the correct values are uncertain due to the lack of $P_h$ information. 

\begin{figure}
\includegraphics[width=0.86\columnwidth]{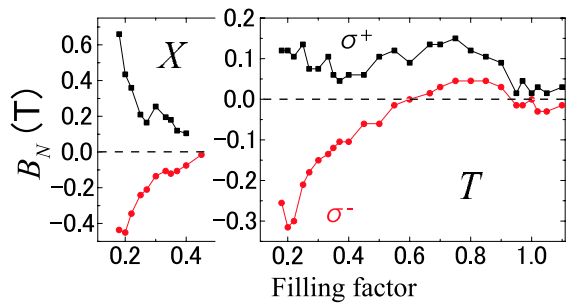}%
\caption{(Color online) The $\nu$-dependence of $B_N$ along the PL peak positions in Fig.~\ref{fig:2Dmap6.eps}. 
\label{fig:FillingDepAlong5.eps}}
\end{figure}

To consider how $\langle S_z \rangle_{\rm{eq}}$ affects $B_N$, 
we should exclude the pronounced difference of $\langle S_z \rangle$ 
between $X$ and $T$ resonant excitations. 
To this end, we extract the values of $B_N$ at the PL peak positions from Fig.~\ref{fig:2Dmap6.eps}.  
The data are shown in Fig.~\ref{fig:FillingDepAlong5.eps} \cite{contTT}. 
The slight $B_N$ increase in the $\nu$-range from 0.4 to 0.8 can be understood from  
the obtained trend of $\langle S_z \rangle_{\rm{eq}}$, which is an almost monotonic increase.
In $\nu <0.4$, $B_N$ does not obey the trend of $\langle S_z \rangle_{\rm{eq}}$. 
The lower the $n_s$, the higher the nuclear polarization obtained. 
This is attributed to the increase of $\langle S_z \rangle$ for both $X$ and $T$ excitations. 
Increasing the number of the injected electron relative to $n_s$ can enhance $\langle S_z \rangle$.
Although a theoretical study on the electron spin pumping in the quantum Hall regime 
is required for complete explanation,  
it should be noted that 
diminishing the doping enhances $\langle S_z \rangle$ in $B$ parallel to the well \cite{EPTX}.

\begin{figure}
\includegraphics[width=0.68\columnwidth]{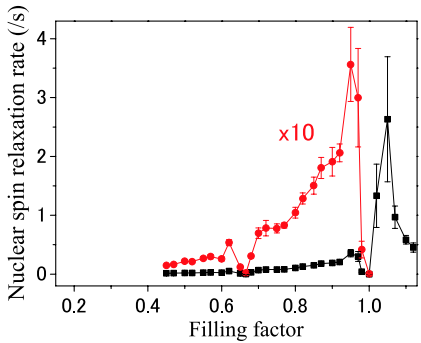}%
\caption{(Color online) 
The $\nu$-dependence of the nuclear spin relaxation rate. 
The red curve shows the magnification of $\nu \le 1$.
\label{fig:NuclearSpinRelaxation.eps}}
\end{figure}

Finally, we consider the optical nuclear polarization in region (III).
In this region, the skyrmion exists under our experimental conditions. 
The low-frequency spin fluctuations associated with the skyrmion destroy the nuclear polarization.
We measured the nuclear spin relaxation  
by changing the waiting time at temporal $\nu$ after optical pumping. 
The time decay of $B_N$ is fitted by the exponential function. 
The observed nuclear spin relaxation rates $1/T_{1N}$ are displayed in Fig.~\ref{fig:NuclearSpinRelaxation.eps} \cite{CommentSkyrmion}. 
The relatively small values of $1/T_{1N}$ at $\nu=2/3$ and $1$ 
are due to the energy gap of the quantum Hall state.
We clearly observe the strong nuclear spin relaxation around $\nu=1$ \cite{Note2}. 
This diminishes the nuclear spin polarization. 
At $\nu=1$, the up spin sublevel of the lowest Landau level is expected to be fully occupied. 
Therefore, the up spin cannot be excited and the photo-excited down spin cannot relax to the up spin. 
This can inhibit the nuclear spin polarization. 

In conclusion, 
we studied nuclear spin polarization in the quantum Hall regime 
through the optically pumped electron spin polarization in the lowest Landau level. 
We found the obvious $\nu$ dependence of the optically induced $B_N$. 
To understand this behavior, 
we constructed a novel estimation method of $\langle S_z \rangle_{\rm{eq}}$
from the photoluminescence polarization.  
This method is based on the fact that $B_N$ is proportional to 
the electron spin polarization difference between the optical pumping and equilibrium conditions. 
On the basis of this estimation method, we obtained the trend of $\langle S_z \rangle_{\rm{eq}}$ 
and thus analyzed the experimental data. 
Finally, we concluded that 
$\nu$ dependence of $B_N$ is understood by 
not fractional quantum Hall states 
but the effect of electron spin polarization through excitons and trions.
The obtained understanding of the optical nuclear spin polarization 
leads to nuclear spins being effectively manipulated by 
combining optical and electrical means. 

\begin{acknowledgments}
The authors are grateful to M. Ohzu, G. Yusa, P. Hawrylak, N. Shibata and J. Hayakawa 
for their fruitful discussions, 
and K. Muraki for providing high-quality wafers. 
One of the authors (TY) was supported by the JSPS Research Fellowship for Young Scientists (No.24-1112). 
\end{acknowledgments}


\begin{thebibliography}{99}
\bibitem{ReasonLP}
To avoid the optical nuclear polarization, we used the linearly polarized laser excitation.
\bibitem{TTcontPL}
The contribution of the triplet trion cannot be neglected in the PL experiments.
\bibitem{NoteA}
In Ref.~\cite{Groshaus}, when the sample was cooled down to 70~mK, 
the $T$ absorption in $\sigma^{-}$ disappeared at $\nu=1/3$ 
due to the full polarization of electron spins. 
In our experiment, the large laser power increases the electron temperature 
and the $T$ transition strength always exists.
\bibitem{NoteStrong}
Although the PL data were acquired under the strong photo-excitation, 
the optical nuclear polarization, which modifies the electron spin polarization,  
was not observed. This is because we used the linearly polarized excitation with high energy for PL.
\bibitem{contTT}
Since the triplet trion transition is merged into the neutral exciton transition, 
there is a small contribution of the triplet trion to $B_N$ in the left side panel.
\bibitem{CommentSkyrmion}
The similar behavior has been reported so far \cite{SpinPhysics}. However, the size of the skyrmion depends on the experimental situation. This measurement in our experiment is valuable.
\end{thebibliography}

\clearpage
\widetext
\setcounter{page}{1}
\setcounter{figure}{0}
\setcounter{equation}{0}

\renewcommand{\thepage}{S\arabic{page}}
\renewcommand{\thefigure}{S\arabic{figure}}
\renewcommand{\theequation}{S\arabic{equation}}


\begin{center}

{\Large
Supplemental Material for

\vspace{0.5cm}

\baselineskip 7mm
Optically Induced Nuclear Spin Polarization in the Quantum Hall regime: \\
The effect of Electron Spin Polarization through Exciton and Trion Excitations
}

\vspace{0.5cm}

by

\vspace{0.3cm}

K. Akiba, S. Kanasugi, T. Yuge, K. Nagase and Y. Hirayama 
\end{center}

\vspace{0.3cm}

In this supplemental material, we show schematics of the experimental setup and the sample 
and we describe the details of our estimation of electron spin polarization 
from the singlet trion photoluminescence (PL) polarization. 

\section{The details of the experimental setup}
\begin{figure}[h]
\begin{center}
\includegraphics[width=0.9\columnwidth]{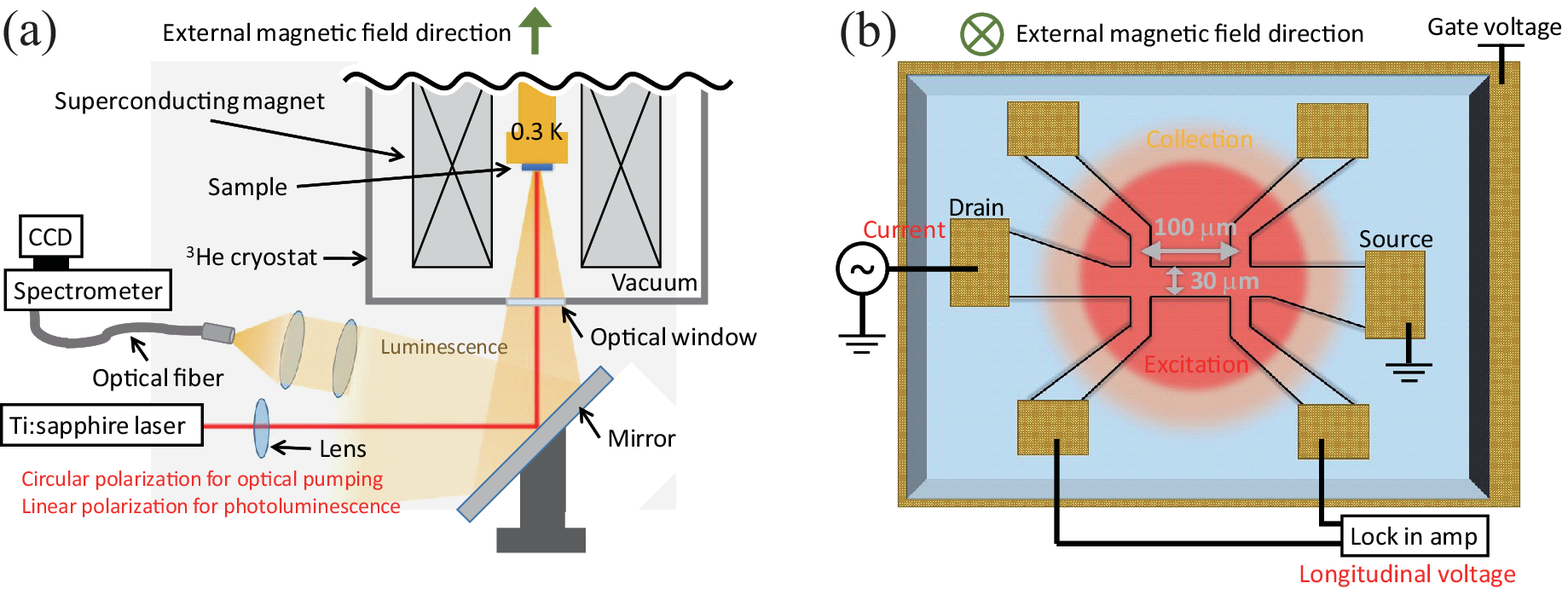}
\end{center}
\caption{(Color online) The schematics of (a) the experimental setup and (b) the sample.} 
\label{fig:detail.eps}
\end{figure}

Figure~\ref{fig:detail.eps} shows the details of our experimental setup. 
The optical pumping and the laser excitation for PL were performed using a Ti:sapphire laser. 
The polarization of light was circular for the optical pumping and linear for the PL. 
The laser beam was focused on the sample 
through an optical window on the bottom of the $^3$He cryostat. 
The temperature of the sample was 0.3~K.  
The direction of the laser beam was parallel to the external magnetic field, 
which was applied perpendicular to the quantum well. 
The beam diameter was $\sim 230\mu$m, which was sufficiently larger than the Hall bar structure, 
and the luminescence collection area covered the laser excitation area, 
as shown in (b). 
The PL was collected through the same optical window (see (a)).  
The collected PL was delivered to the spectrometer by using an optical fiber 
and the spatially split spectrum was recorded by a CCD camera. 
The 2-dimensional electron system of the Hall bar was electrically contacted with 6 electrodes, 
which are shown in (b). 
These contacts enable us to measure the resistance by the 4-terminal method 
and we tuned the electron density in the quantum well by applying a voltage to the back gate. 

\section{The details of the estimation of electron spin polarization from the singlet trion photoluminescence polarization}
We construct the relationship between the electron spin polarization and the PL polarization
by modeling the oscillator strength (OS) of a singlet trion 
as proportional to the number of unpaired electrons with the spin opposite to the photo-created electron spin. 
By using the constructed relationship, 
we show an example of the estimation of electron spin polarization 
from the experimentally obtained PL polarization. 

\subsection{Singlet trion photoluminescence polarization}
A singlet trion is two conduction band electrons with opposite spin 
coupling with a valence band hole due to the Coulomb interaction. 
The PL of the singlet trion 
shows right or left circular polarization ($\sigma^{+}$ or $\sigma^{-}$)
depending on the direction of the hole spin. 
Since the PL intensity is generally proportional to 
the number of the photo-excited particles and their OSs, 
the $\sigma^{+}$ and $\sigma^{-}$ PL intensity are given by 
\begin{align}
I_{\sigma^{+}}=c\,N_{\Uparrow}/\tau_{\Uparrow},\quad 
I_{\sigma^{-}}=c\,N_{\Downarrow}/\tau_{\Downarrow}, \label{Ipm}
\end{align}
where $c$ is a constant, 
$N_{\Uparrow (\Downarrow)}$ is the number of singlet trion 
with the up (down) spin $\Uparrow (\Downarrow)$ hole, 
and $1/\tau_{\Uparrow (\Downarrow)}$ is its OS. 
We define the PL polarization $P_L$ for singlet trion as 
\begin{align}
P_L=\frac{I_{\sigma^{+}}-I_{\sigma^{-}}}{I_{\sigma^{+}}+I_{\sigma^{-}}}. \label{Ppl}
\end{align}
Note that the energies of $\sigma^{+}$ and $\sigma^{-}$ PL are different.

According to the study of singlet trion absorption \cite{Groshaus}, 
since the photo-created electron needs to pair with an opposite spin, 
the OS of each trion is proportional to the number of unparied electrons with opposite spin. 
Thus, we obtain
\begin{align}
1/\tau_{\Uparrow}=Cf_{\uparrow}N_{\uparrow}, \quad 
1/\tau_{\Downarrow}=Cf_{\downarrow}N_{\downarrow}, \label{OSpm}
\end{align}
where $C$ is a constant, 
$f_{\uparrow (\downarrow)}$ is the fraction of spin up (down) $\uparrow (\downarrow)$ electrons 
that are unpaired, 
and $N_{\uparrow (\downarrow)}$ is the number of $\uparrow (\downarrow)$ electrons. 

Assuming $f_{\uparrow (\downarrow)}=1$ for $\nu \le 1$ \cite{Groshaus}, 
we substitute Eqs.~(\ref{Ipm}) and (\ref{OSpm}) in Eq.~(\ref{Ppl}) and obtain 
\begin{align}
P_L=\frac{P_h+P_e}{1+P_e\,P_h}, \label{Pple}
\end{align}
where $P_h=(N_{\Uparrow}-N_{\Downarrow})/(N_{\Uparrow}+N_{\Downarrow})$ 
is the trion spin polarization
and $P_e=(N_{\uparrow}-N_{\downarrow})/(N_{\uparrow}+N_{\downarrow})$ 
is the electron spin polarization. 
The trion spin polarization $P_h$ is identical to the hole spin polarization, 
because the two electrons in the trion form a spin singlet state, which has zero resultant spin. 

Accordingly, Eq.~(\ref{Pple}) can be transformed into 
\begin{align}
P_e=\frac{P_L-P_h}{1-P_L\,P_h}, \label{Pe1}
\end{align}
and we can estimate $P_e$ from $P_L$ when we attain the information of $P_h$. 

For $\nu >1$, we assume $f_{\uparrow (\downarrow)}=1-[(N_{\uparrow}+N_{\downarrow})-N_\phi]/N_{\uparrow(\downarrow)}$, 
where $N_\phi$ is the degeneracy factor of the spin split Landau level \cite{Groshaus}.  
The similar calculation with $(N_{\uparrow}+N_{\downarrow})/N_\phi=\nu$ gives
\begin{align}
P_e=\frac{P_L-P_h}{1-P_L\,P_h}\frac{2-\nu}{\nu}, \label{Pe2}
\end{align}
where $\nu$ is filling factor.

\subsection{Estimation of electron spin polarization}
We measured the PL under experimental conditions almost the same as those in the main manuscript.
Therefore, the sample we used was a single 18-nm GaAs/Al$_{0.33}$Ga$_{0.67}$As quantum well 
that was cooled in a cryogen free $^3$He refrigerator down to 0.3 K. 
The linearly polarized excitation laser beam with photon energy of $1.58$~eV was injected 
through an optical window on the bottom of the cryostat 
and the left circularly polarized PL was collected 
through the same window with a spectral resolution better than 0.2~meV 
under the external magnetic field $B$ of $\pm$7.15~T. 
Negative $B$ was used to 
avoid the optical loss difference caused by the different optical system. 
The left circular polarization under negative $B$ corresponds to 
the right circular polarization under positive $B$. 
The difference from the main manuscript is the much smaller laser power of 2.4~mW/cm$^2$.

\begin{figure}[h]
\begin{center}
\includegraphics[width=65mm]{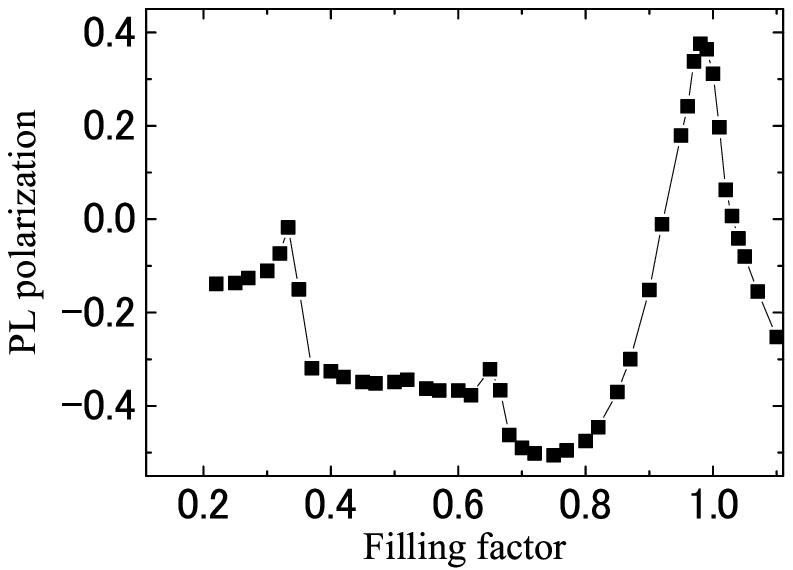}
\end{center}
\caption{The $\nu$-dependence of the singlet trion PL polarization obtained from the PL measurement with the excitation laser power of 2.4~mW/cm$^2$.} 
\label{fig:PLpol2uW.eps}
\end{figure}

We recorded the dependence of the PL on $\nu$ by using the back gate. 
The obtained singlet trion PL polarization $P_L$ from Eq.~(\ref{Ppl}) is shown in Fig.~\ref{fig:PLpol2uW.eps}. 
We clearly observe the peak at $\nu=1/3,\ 2/3,\ $and 1. 
In the main manuscript, we only observe the peak at $\nu=1$ in Fig.~3 (a). 
This difference can be understood by the weaker laser heating effect 
because only the excitation laser power is different between Fig.~\ref{fig:PLpol2uW.eps} and Fig.~3 (a). 

\begin{figure}[h]
\begin{center}
\includegraphics[width=65mm]{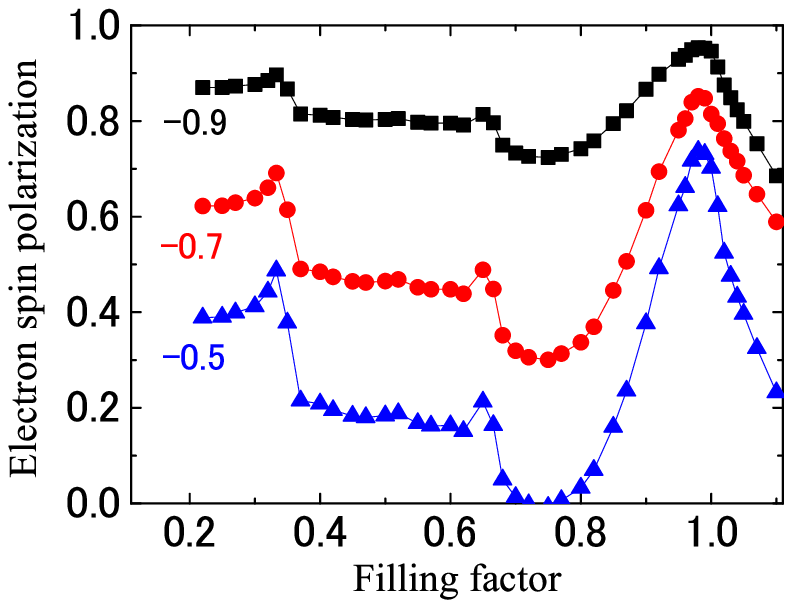}
\end{center}
\caption{(Color online) The $P_e$ calculated with $P_h=-0.9, -0.7$ and $-0.5$ 
using Eqs.~(\ref{Pe1}) and (\ref{Pe2}).} 
\label{fig:epol2uW.eps}
\end{figure}

To calculate $P_e$ from Eqs.~(\ref{Pe1}) and (\ref{Pe2}), we have to acquire $P_h$. 
The value of $P_h$ generally depends on $B$ \cite{Kukushkin}. 
In our experiment, $\nu$ is tuned by using the back gate and the $B$ is fixed. 
Therefore, $P_h$ is expected to be constant in all ranges of $\nu$. 
If one has the knowledge of a certain value of $\nu$ with $P_e=0$ (e.g. $\nu=2$), 
$P_L$ is equal to $P_h$ and one can obtain the value of $P_h$ from the PL polarization experiment. 
However, we cannot extract the correct value of $P_h$ from our experimental results. 
The fact that the g-factor of GaAs is negative 
indicates that $P_e$ ranges from 0 to 1 in quantum Hall regime. 
We assume the constant value of $P_h$ is given by 
the resultant value of $P_e$ satisfied with this range. 
Figure~\ref{fig:epol2uW.eps} shows the $P_e$ calculated with $P_h=-0.9, -0.7$ and $-0.5$ 
using Eqs.~(\ref{Pe1}) and (\ref{Pe2}). 
The obtained behavior of $P_e$ is consistent with the previous studies \cite{Groshaus, KukushkinSpin}
This can validate our estimation method. 

Although we cannot obtain the correct value of $P_e$ 
due to the lack of information of $P_h$ in our experiments, 
we can at least conclude that the trend of $P_e$ reflects that of $P_L$ under constant $B$ 
from the comparison between Figs.~\ref{fig:PLpol2uW.eps} and \ref{fig:epol2uW.eps}. 
Furthermore, once one obtains the correct value of $P_h$ or 
determines the $P_e$ value at a certain $\nu$ experimentally 
(e.g., from the nuclear magnetic resonance), 
the estimation of $P_e$ from $P_L$ we constructed here becomes a fairly useful method.


\end{document}